\documentclass[journal=nalefd,manuscript=letter,layout=twocolumn]{achemso}
\usepackage[version=3]{mhchem} 
\usepackage{graphicx}
\usepackage{color}
\usepackage{amssymb,amsmath}
\usepackage{bm,achemso}
\usepackage{hyperref}
\usepackage{url}

\newcommand{\BSCCO}{Bi$_2$Sr$_2$CaCu$_2$O$_{8+x}$}

\let\oldmaketitle\maketitle
\let\maketitle\relax

\author{Ilija Zeljkovic}
\email{ilija.zeljkovic@bc.edu}
\affiliation{Department of Physics, Harvard University, Cambridge, MA 02138, U.S.A.}
\author{Jouko Nieminen}
\affiliation{Tampere University of Technology, Finland}
\author{Dennis Huang}
\affiliation{Department of Physics, Harvard University, Cambridge, MA 02138, U.S.A.}
\author{Tay-Rong Chang}
\affiliation{Department of Physics, National Tsing Hua University, Hsinchu 30013, Taiwan}
\author{Yang He}
\affiliation{Department of Physics, Harvard University, Cambridge, MA 02138, U.S.A.}
\author{Horng-Tay Jeng}
\affiliation{Department of Physics, National Tsing Hua University, Hsinchu 30013, Taiwan}
\alsoaffiliation{Institute of Physics, Academia Sinica, Taipei 11529, Taiwan}
\author{Zhijun Xu}
\author{Jinsheng Wen}
\author{Genda Gu}
\affiliation{Brookhaven National Laboratory, Upton, NY 11973, U.S.A.}
\author{Hsin Lin}
\author{Robert S. Markiewicz}
\author{Arun Bansil}
\affiliation{Department of Physics, Northeastern University, Boston, MA 02115, U.S.A.}
\author{Jennifer E. Hoffman}
\email{jhoffman@physics.harvard.edu}
\affiliation{Department of Physics, Harvard University, Cambridge, MA 02138, U.S.A.}

\title{Nanoscale interplay of strain and doping in a high-temperature superconductor}

\begin{document}

\twocolumn[
\begin{@twocolumnfalse}
\oldmaketitle
\begin{abstract}
The highest temperature superconductors are electronically inhomogeneous at the nanoscale, suggesting the existence of a local variable which could be harnessed to enhance the superconducting pairing. Here we report the relationship between local doping and local strain in the cuprate superconductor \BSCCO. We use scanning tunneling microscopy to discover that the crucial oxygen dopants are periodically distributed, in correlation with local strain. Our picoscale investigation of the intra-unit-cell positions of all oxygen dopants provides essential structural input for a complete microscopic theory.
\end{abstract}
\end{@twocolumnfalse}
]

Cuprate superconductors display startling nanoscale inhomogeneity in essential properties such as the spectral gap $\Delta$,\cite{HowaldPRB2001, PanNature2001, LangNature2002} collective mode energy $\Omega$,\cite{LeeNature2006} and even the pairing temperature $T_p$.\cite{GomesNature2007} The identity of the primary local variable controlling this electronic inhomogeneity has been debated for more than a decade. Both dopants\cite{ZeljkovicPCCP2013} and strain\cite{SlezakPNAS2008} have been empirically linked with electronic structure. Theoretical models have argued for the primacy of charge,\cite{MartinPhysicaC2001} strain,\cite{NunnerPRL2005} or a carefully tuned combination of both.\cite{ChenNJP2011} A set of strain theories have specifically explored the relationship between the apical oxygen height and the superconducting pairing strength, predicting both enhancement\cite{OhtaPRB1991, RaimondiPRB1996, PavariniPRL2001, BergmanMaterials2011} and reduction\cite{MoriPRL2008} of pairing strength with increasing apical oxygen height. However, microscopic theoretical understanding of cause and effect has been stalled by uncertainty about the precise dopant locations.

\begin{figure}[!t]
\includegraphics[width=1.0\columnwidth,clip]{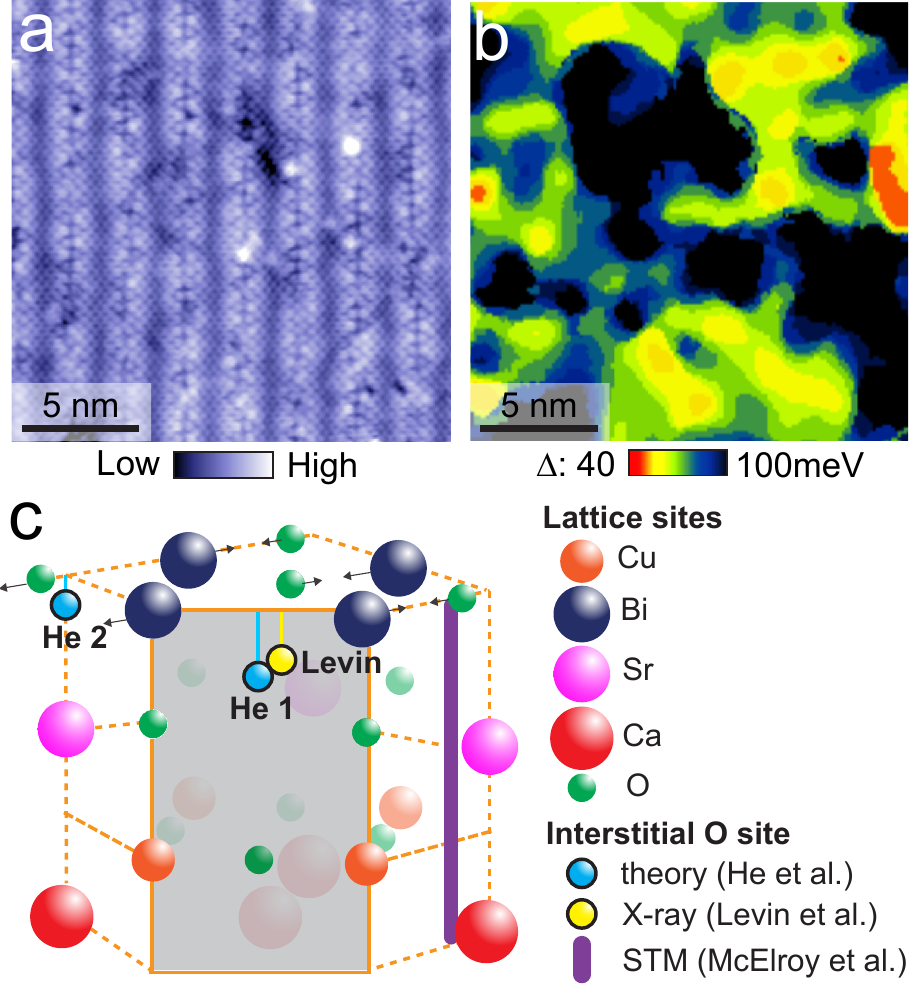}
\caption{(a) Typical STM topograph of the BiO cleaved surface of underdoped Bi2212 with $T_c$=55 K, acquired at 150 pA, +1 V and 6 K. (b) Two-dimensional map of the spectral gap magnitude $\Delta$ acquired over the region of the sample shown in (a), depicting the nanoscale electronic inhomogeneity present in this family of materials. (c) Schematic representation of the top four layers (BiO, SrO, CuO$_2$ and Ca) of the Bi2212 unit cell. Gray shaded area represents a vertical cut through the crystal structure to emphasize possible positions of $\mathrm{O_i(Bi)}$. Light blue spheres represent the $\mathrm{O_i(Bi)}$ positions predicted by theory \cite{HePRL2006}, and yellow sphere is the position extracted from X-ray experiments \cite{LevinJPCM1994}. Purple vertical line shows the position of $\mathrm{O_i(Bi)}$ obtained by previous STM experiments which did not take into account the orthorhombic structural distortion \cite{McElroyScience2005}. Black arrows denote the direction of orthorhombic distortion of Bi and O atoms in the BiO layer.
\label{fig:CrystalStructure}}
\end{figure}

\BSCCO\ (Bi2212) presents an excellent test case to address these questions experimentally, as it harbors several intrinsic sources of strain and doping whose local relation to the electronic structure can be measured. The largest source of strain in Bi2212 is an incommensurate structural buckling known as the ``supermodulation'' with a period of $\sim$26 \AA, oriented at 45$^{\circ}$ from the Cu-O bond \cite{PetricekPRB1990, LePagePRB1989} (Fig.~\ref{fig:CrystalStructure}a). A second source of strain is a commensurate orthorhombic distortion which shifts two sub-lattices in opposite directions, perpendicular to the supermodulation wavevector, primarily in the BiO plane \cite{SubramanianScience1988, ZeljkovicNatMat2012}. Both modulations distort lattice oxygen atoms from their ideal positions by more than 0.5 \AA. Superconductivity in Bi2212 is induced by two types of interstitial oxygen dopants, each of which are thought to donate up to two holes to the CuO$_2$ plane: ``type-A'' (which are likely to be in the SrO layer, so we refer to them hereafter as $\mathrm{O_i(Sr)}$) \cite{ZhouPRL2007, ZeljkovicScience2012} and ``type-B'' (which are likely to be in the BiO layer, so we refer to them hereafter as $\mathrm{O_i(Bi)}$) \cite{McElroyScience2005}. Underdoped Bi2212 also contains a significant number of apical oxygen vacancies (AOVs), which are thought to donate electrons to the CuO$_2$ plane \cite{ZeljkovicScience2012}.

McElroy \emph{et al.}\ initially observed a correlation between the positions of $\mathrm{O_i(Bi)}$ (hole donors) and regions of enhanced spectral gap $\Delta$ \cite{McElroyScience2005} (for example, the dark regions in Fig.~\ref{fig:CrystalStructure}b). This finding contrasted with expectations from the global trend which associated increased hole doping with reduced $\Delta$ \cite{MiyakawaPRL1998}. It suggested that the dopant's local charging effect was insignificant, in comparison to its induced local strain. Slezak \emph{et al.}\ then claimed to definitively isolate strain as the controlling variable, with the observation that $\Delta$ is $\sim$10\% larger at the peaks than at the troughs of supermodulation \cite{SlezakPNAS2008}.

Although the first generation of STM experiments \cite{McElroyScience2005, SlezakPNAS2008} and theories \cite{NunnerPRL2005, AndersenPRB2007, MoriPRL2008} on Bi2212 seemed to support the role of strain as the primary local variable controlling $\Delta$, these studies missed several key aspects of the problem. First, failure to include the orthorhombic distortion in the relevant BiO dopant plane cast some doubt on the detailed microscopic models. Second, and more importantly, neither theory nor experiment took into account $\mathrm{O_i(Sr)}$ or AOVs, which have been discovered only recently.\cite{ZeljkovicScience2012} Arguments for the primacy of strain hinged on the assumption that $\mathrm{O_i(Bi)}$ oxygens were the sole dopants in the system.

Here we use a home-built, low-temperature, picoscale STM to perform a detailed study of the interplay of both types of strain with all three types of oxygen defects in Bi2212, and we present two major advances. First, we report the intra-unit-cell locations of all interstitial dopants, resolving discrepant results in the literature and providing a reliable structural basis for microscopic models. Second, we reveal the relationship of the newly identified $\mathrm{O_i(Sr)}$ and AOVs to the supermodulation, suggesting that the previously reported correlation between $\Delta$ and the supermodulation \cite{SlezakPNAS2008} may in fact be directly caused by the periodically placed dopants rather than the strain from the supermodulation itself.


We first address conflicting reports of the intra-unit-cell location of $\mathrm{O_i(Bi)}$. X-ray studies on Bi2212 \cite{LePagePRB1989, YamamotoPRB1990, LevinJPCM1994, PetricekPRB1990} located interstitial oxygen dopants in the BiO plane, and Levin \emph{et al.}\ reported their lateral location halfway between neighboring Bi atoms \cite{LevinJPCM1994} (yellow sphere in Fig.\ \ref{fig:CrystalStructure}c). Direct imaging by McElroy \emph{et al.}\ showed a different lateral position (purple line in Fig.\ \ref{fig:CrystalStructure}c indicates the insensitivity of STM to $c$-axis position). Finally He \emph{et al.}\ used density functional theory (DFT) to determine two different stable positions of interstitial O atoms (light blue spheres in Fig.\ \ref{fig:CrystalStructure}c, with ``He1'' being more energetically favorable than ``He2'' \cite{HePRL2006}). However, none of these conflicting efforts took into account the orthorhombic structural distortion, which displaces lattice oxygens laterally by 0.55 \AA\ in the BiO layer.\cite{MilesPhysicaC1998} 

\begin{figure*}[!t]
\includegraphics[width=2\columnwidth,clip]{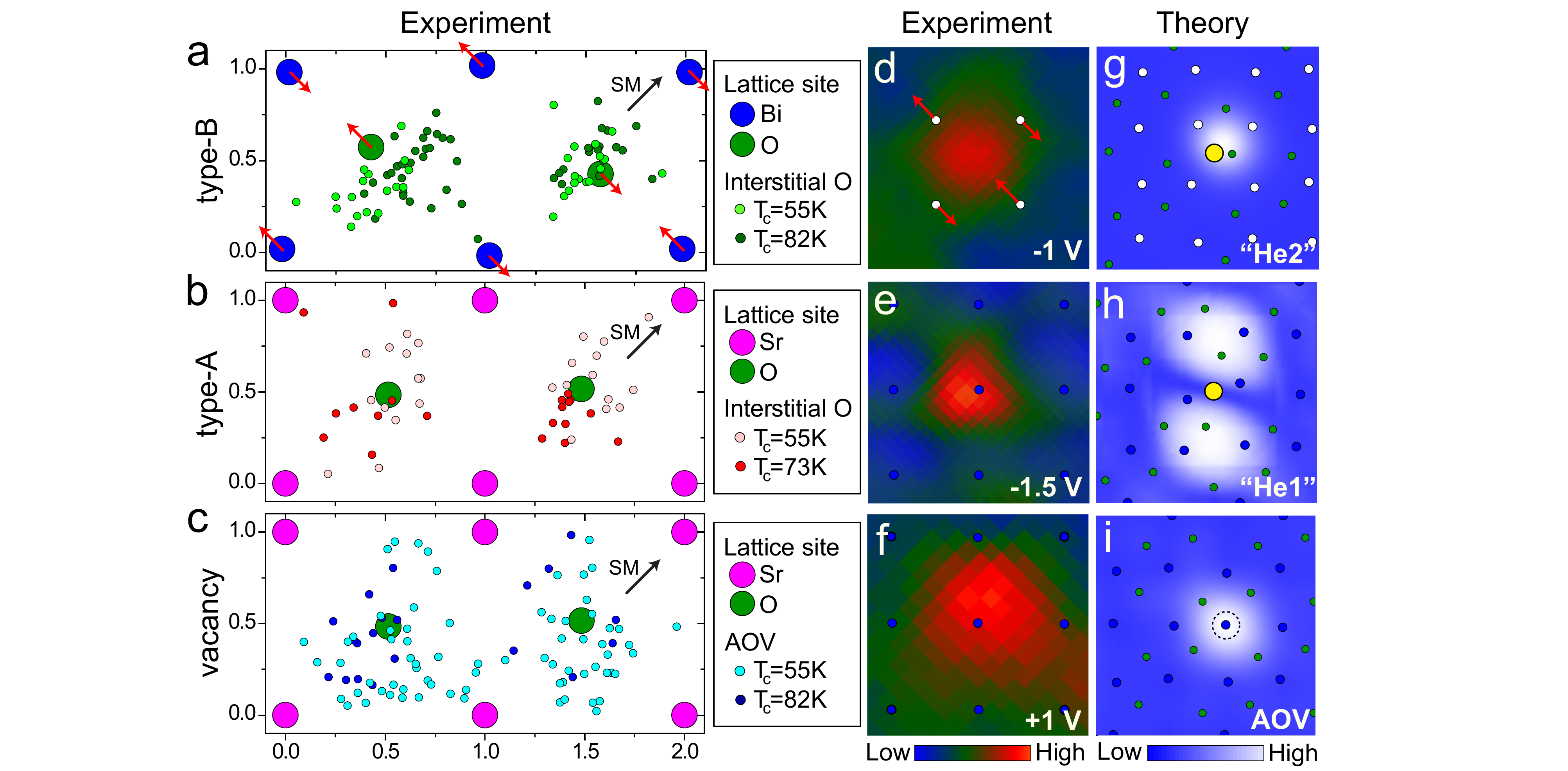}
\caption{(a-c) Distributions of $\mathrm{O_i(Bi)}$, $\mathrm{O_i(Sr)}$ and AOVs within the orthorhombic unit cell for several areas within two samples. Coordinates of the lattice atoms have been taken from neutron diffraction measurements \cite{MilesPhysicaC1998}, and are expressed in units of $a_0$=3.83 \AA. Red arrows denote the direction of the orthorhombic shift, and black arrows show the wavevector direction of the supermodulation (SM). (d-f) 1 nm $dI/dV$ maps containing a single $\mathrm{O_i(Bi)}$ (-1 V), $\mathrm{O_i(Sr)}$ (-1.5 V) and AOV (+1 V) respectively, for the $T_c$=73K sample. Setup conditions are -1 V, -1.5 V, +1 V and 300 pA, 100 pA, 200 pA respectively. White circles show the idealized Bi lattice from the simultaneously acquired and drift-corrected BiO topographs \cite{LawlerNature2010}. Simulations of 1.5 nm constant height $dI/dV$ images showing a single interstitial O at -1 V at (g) ``He2'' position and (h) ``He1'' position, as well as a single AOV at +1 V (i). White, green and yellow circles in (g-i) represent positions of lattice Bi atoms, lattice oxygen atoms, and interstitial oxygen dopants, respectively. Dashed circle in (i) denotes the AOV position. 
\label{fig:UCPositions}}
\end{figure*}

We acquire STM topographs and simultaneous $dI/dV$ images of the BiO layer at +1 V, -1 V, and -1.5 V sample bias, and locate the exact positions of three types of oxygen defects with respect to the Bi lattice seen in the topographs (detailed description given in Supporting Information). Figure \ref{fig:UCPositions}a shows a scatter plot of $\mathrm{O_i(Bi)}$ locations within the full orthorhombic unit cell of the BiO layer. In a perfect tetragonal cell, the average lateral location of $\mathrm{O_i(Bi)}$ would appear to coincide with the lattice O position in the BiO layer (O(Bi)), which seems impossible due to the lattice O already there. Here we resolve the conflict by mapping the distribution of $\mathrm{O_i(Bi)}$ locations into the two distinct halves of the orthorhombic unit cell. Each lattice O shifts 0.55 \AA\ (15\% of the tetragonal cell) away from the high symmetry point \cite{MilesPhysicaC1998}, and we now find that $\mathrm{O_i(Bi)}$ are located on the opposite side of the unit cell from the lattice oxygen, consistent with the second most energetically favorable position ``He2'', shown in Fig.\ \ref{fig:CrystalStructure}c. Thus the spatial distribution of $\mathrm{O_i(Bi)}$ is directly connected to the orthorhombic strain, as well as ``stretched'' along one Bi-O direction parallel to the supermodulation wavevector in both samples studied in detail (Fig.\ \ref{fig:UCPositions}a).

To support our empirical identification of $\mathrm{O_i(Bi)}$, we performed local density approximation (LDA) calculations to optimize the structure in the case of an orthorhombic unit cell starting from ``He1'' and ``He2'' configurations. Our calculations are augmented by Green's function based simulation of the $dI/dV$ maps (see Supporting Information) which are shown in Figs.\ \ref{fig:UCPositions}g and \ \ref{fig:UCPositions}h for ``He2'' and ``He1'', respectively. The simulated $dI/dV$ map for an interstitial oxygen dopant close to the ``He2'' position correctly reproduces the experimental $dI/dV$ map (Fig.\ \ref{fig:UCPositions}g). In contrast, the dominant tunneling paths for the ``He1'' interstitial position pass through the adjacent Bi atoms, with destructive interference at the O(Bi) directly above. The resulting $dI/dV$ simulation has a two-lobe structure (Fig.\ \ref{fig:UCPositions}h) inconsistent with the experimental observation in Fig.\ \ref{fig:UCPositions}d.

We use the same procedure to determine the intra-unit-cell positions of $\mathrm{O_i(Sr)}$ interstitial oxygen dopants. Since these dopants were predicted to occur closer to, or even possibly within, the SrO plane \cite{ZhouPRL2007}, we portray the scatter plot of their positions within the full orthorhombic SrO unit cell. Figure \ref{fig:UCPositions}b shows the $\mathrm{O_i(Sr)}$ locations centered around the lattice O site in the SrO layer (O(Sr)). No difference can be seen between the distributions in the two orthorhombic unit cells, which is not surprising, because the orthorhombic distortion affecting O(Sr) is less than 0.8\% of the tetragonal cell, much smaller than the corresponding distortion of O(Bi). However, O(Sr) is vertically displaced by $\sim$0.7\AA\ above the horizontal plane containing the Sr atoms \cite{MilesPhysicaC1998}, which may leave enough space for $\mathrm{O_i(Sr)}$ to position themselves just below O(Sr). The surprising conclusion of our direct imaging experiments is that both types of interstitial oxygen dopants occupy positions inconsistent with the theoretically predicted position ``He1'' and the X-ray position ``Levin'' shown in Fig.\ \ref{fig:CrystalStructure}c.

For completeness, we plot the AOV distribution in Fig.\ \ref{fig:UCPositions}c. The locations in Fig.\ \ref{fig:UCPositions}c show a greater scatter than in Figs.\ \ref{fig:UCPositions}a-b, which may be due to the apparent larger size of the AOVs, compared to interstitial oxygen dopants. To illustrate the size difference between the 3 dopant signatures, Figs.\ \ref{fig:UCPositions}d-f show 1 nm $dI/dV$ maps with one $\mathrm{O_i(Bi)}$, one $\mathrm{O_i(Sr)}$ and one AOV, respectively. The $dI/dV$ simulation for the AOV is in good agreement with the experiment (Fig.\ \ref{fig:UCPositions}i).

\begin{figure}[!t]
\includegraphics[width=1.0\columnwidth,clip]{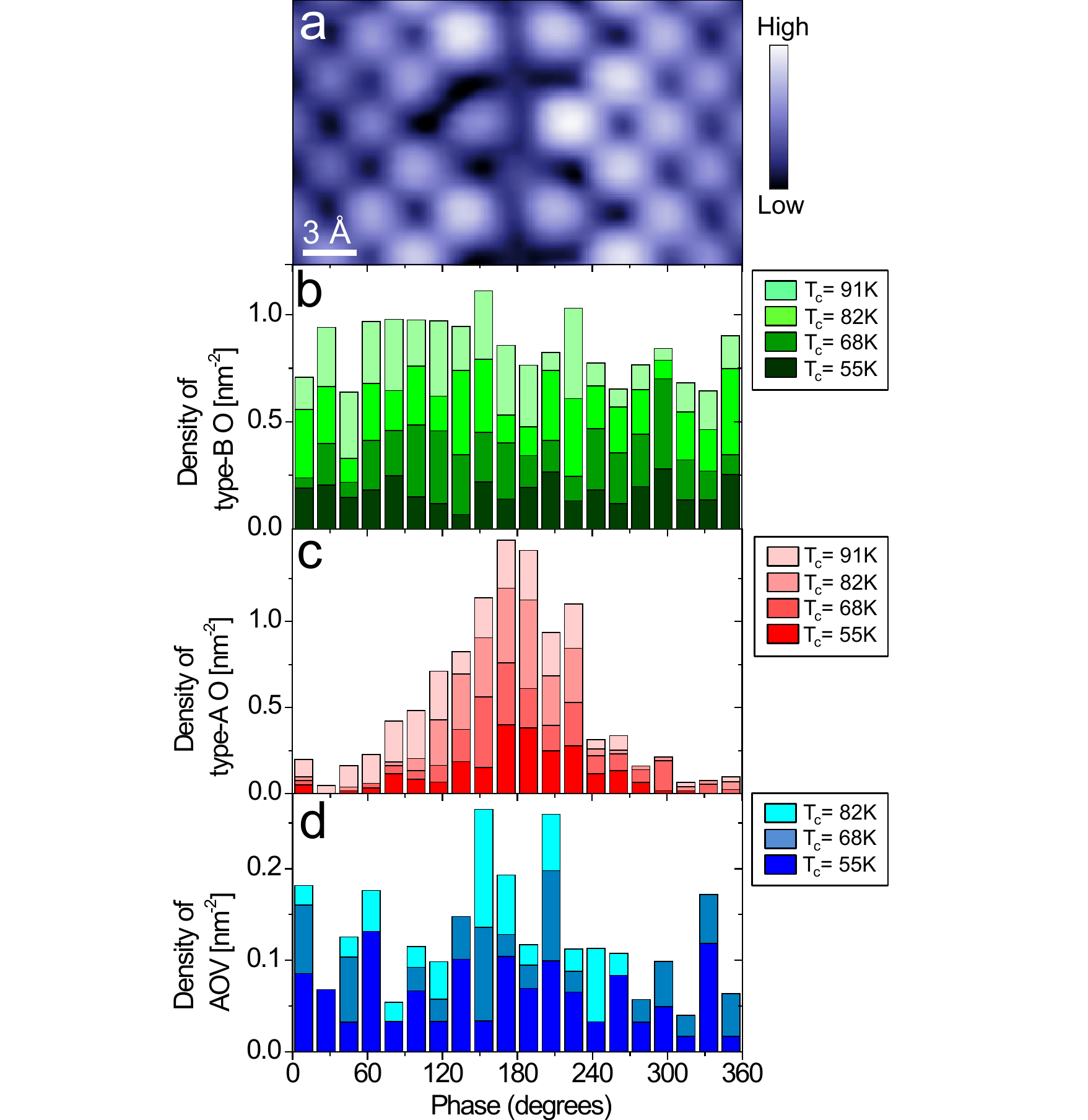}
\caption{(a) BiO layer STM topograph acquired at 200 mV and 50 pA, $\sim$2.5 nm wide, showing distortions in the Bi lattice over one period of the supermodulation. (b-d) Histograms of densities of $\mathrm{O_i(Bi)}$, $\mathrm{O_i(Sr)}$, and AOVs respectively as a function of the phase of the supermodulation. Different shades of red, green and blue indicate data obtained on samples of different doping concentrations (size of square regions used vary across different dopings from $\sim$28 nm to $\sim$35 nm). The crest of the supermodulation, thought to correspond to the minimum in apical oxygen height\cite{YamamotoPRB1990}, is at 180$^{\circ}$ as emphasized by the topograph in (a).
\label{fig:SMDistribution}}
\end{figure}

Our second main result addresses the question of whether the previously observed correlation between the supermodulation and $\Delta$ arises from strain alone. \cite{SlezakPNAS2008} We determine the average density of each dopant type as a function of the supermodulation phase, following Slezak's algorithm. \cite{SlezakPNAS2008} Figure \ref{fig:SMDistribution}a shows a 2.5 nm wide atomically-resolved topograph of Bi2212 with the supermodulation crest running vertically down the center. We confirm the lack of correlation between the $\mathrm{O_i(Bi)}$ and the supermodulation \cite{SlezakPNAS2008} (Fig.\ \ref{fig:SMDistribution}b). In contrast, we discover that the $\mathrm{O_i(Sr)}$ are strongly correlated with the crest of the supermodulation for all four samples studied (Fig.\ \ref{fig:SMDistribution}c). We hypothesize that the supermodulation crests create larger inter-atomic spacing and allow $\mathrm{O_i(Sr)}$ to fit there. AOVs also tend to appear at the crest of the supermodulation (Fig.\ \ref{fig:SMDistribution}d) but the smaller overall density of AOVs gives weaker statistical significance to this observation.


\begin{figure}[!t]
\includegraphics[width=1.0\columnwidth]{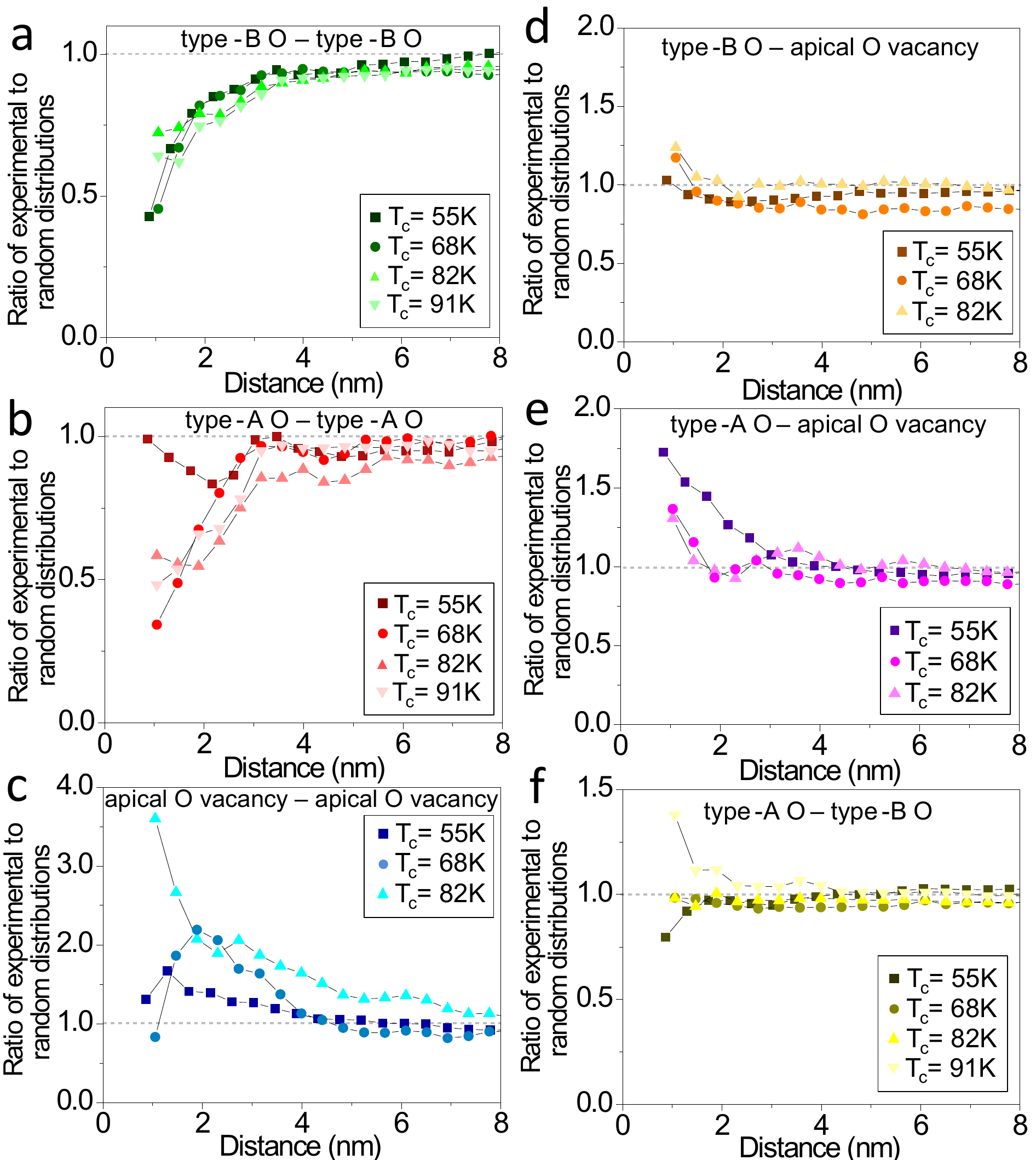}
\caption{Ratios of experimental to random dopant distributions as a function of distance for (a) $\mathrm{O_i(Bi)}$ -- $\mathrm{O_i(Bi)}$, (b) $\mathrm{O_i(Sr)}$ -- $\mathrm{O_i(Sr)}$, (c) AOV -- AOV, (d) $\mathrm{O_i(Bi)}$ -- AOV, (e) $\mathrm{O_i(Sr)}$ -- AOV , and (f) $\mathrm{O_i(Sr)}$ -- $\mathrm{O_i(Bi)}$. Ratios greater than 1.0 indicate inclination of two types of dopants to attract each other compared to a random distribution, whereas ratios less than 1.0 signal tendency of the two types of dopants to repel each other.
\label{fig:Correlations}}
\end{figure}

Finally, we search for clustering and correlations between the three types of oxygen defects. Figure \ref{fig:Correlations} shows the measured likelihood that a dopant of one type ($X$) occurs at a certain distance from another dopant of the same or different type ($Y$), divided by the corresponding likelihood for a simulated random distribution of all dopants (see Supporting Information). Both $\mathrm{O_i(Sr)}$ and $\mathrm{O_i(Bi)}$ show a tendency to repel interstitials of their own species (Figs.\ \ref{fig:Correlations}a-b), consistent with the expectation for particles of like charge. The anomalous distribution of $\mathrm{O_i(Sr)}$ in the $T_c=55$ K sample is likely affected by AOVs, as the two have a strong positive correlation in all samples (Fig.\ \ref{fig:Correlations}e), with the correlation being the strongest in the $T_c=55$ K sample. Indeed, since AOV sites are relatively positively charged, one would expect negatively charged interstitial oxygen atoms to be attracted towards them. $\mathrm{O_i(Bi)}$ are also positively correlated to AOVs (Fig.\ \ref{fig:Correlations}d), but this correlation is much weaker than the correlation between $\mathrm{O_i(Sr)}$ and AOVs. Since AOVs have a well-defined position in the SrO layer, this information supports the hypothesis that $\mathrm{O_i(Sr)}$ are located in the SrO layer \cite{ZhouPRL2007}. Finally, we find that the AOVs tend to cluster in all samples studied (Fig.\ \ref{fig:Correlations}c). The relationship between $\mathrm{O_i(Sr)}$ and $\mathrm{O_i(Bi)}$ is weak and inconsistent between samples (Fig.\ \ref{fig:Correlations}f).

Existing microscopic theory has focused primarily on $\mathrm{O_i(Bi)}$ in the ``He1'' position of a tetragonal BiO lattice, and their effect on $\Delta$ \cite{NunnerPRL2005}. To reconcile the observed local trend ($\mathrm{O_i(Bi)}$ hole donors correlated to regions of large $\Delta$) with the global trend (increased hole doping reduces $\Delta$) required a carefully tuned phenomenological model \cite{ChenNJP2011}. Our work provides a more natural explanation to reconcile the local and global trends. We have found that regions of large $\Delta$ are only weakly correlated to $\mathrm{O_i(Bi)}$, with strongest correlation to AOVs (electron donors) and intermediate correlation to $\mathrm{O_i(Sr)}$ (hole donors) \cite{ZeljkovicScience2012}. Here we show explicitly that AOVs are strongly correlated with $\mathrm{O_i(Sr)}$ (Fig.\ \ref{fig:Correlations}e) and more weakly with $\mathrm{O_i(Bi)}$ (Fig.\ \ref{fig:Correlations}d), so any apparent correlation between $\mathrm{O_i(Sr)}$ or $\mathrm{O_i(Bi)}$ interstitial oxygen hole donors and regions of large $\Delta$, such as that observed at the supermodulation crests (Fig.\ \ref{fig:SMDistribution}c and Ref.\ \cite{SlezakPNAS2008}), may be reconciled as a byproduct of stronger electron donation from the nearby AOVs.


The correlations we report here suggest a simple picture in which $\Delta$ may be locally controlled by dopant charge alone, consistent with global trends. To explain occasional patches of large $\Delta$ which are not coincident with AOVs or supermodulation crests (see Supporting Information), we propose several possibilities. First, excess Bi is typically used to facilitate the Bi2212 growth process, resulting in $\sim$5\% Bi$^{3+}$ substitutions at the Sr$^{2+}$ site \cite{EisakiPRB2004}. Although we have not directly imaged these Sr site defects \cite{KinodaPRB2003}, we expect them to be electron donors, and indeed they have been found to locally increase $\Delta$ \cite{ZeljkovicPCCP2013}. Second, there may be another unknown and rare impurity. Third, electronic interference effects may produce regions of large $\Delta$ where there is no dopant directly present \cite{NunnerPRL2005}. We cannot rule out the possibility of residual strain effects where no dopants are present.


In conclusion, we have used STM experiments, supported by \emph{ab-initio} structural optimization and STM simulation, to locate three species of dopants with picoscale precision in Bi2212.
Our results overturn two long-held beliefs. First, we resolve discrepant reports of the intra-unit-cell position of oxygen dopants in Bi2212 \cite{LevinJPCM1994, McElroyScience2005, HePRL2006}. Second, we find that $\mathrm{O_i(Sr)}$ and apical oxygen vacancies are correlated with the supermodulation, questioning the belief that strain alone controls $\Delta$ \cite{SlezakPNAS2008, AndersenPRB2007, MoriPRL2008}.
Armed with the detailed knowledge of dopant locations and strain in Bi2212, new theoretical models can more accurately compute $\Delta$, collective mode energy $\Omega$, and $T_c$, to address the microscopic pairing mechanism in cuprate superconductors. Furthermore, we note that dramatic changes in $T_c$ are predicted in systems with periodic lines of dopants of varying duty cycle,\cite{GorenPRB2011} and we suggest the potential utility of the supermodulation as a scaffold for such dopant arrangements.

\begin{acknowledgement}
The work at Harvard University was supported by the Air Force Office of Scientific Research under grant FA9550-06-1-0531, and the US National Science Foundation under grant DMR-0847433. The work at Brookhaven National Laboratory was supported by DOE contract at DE-AC02-98CH10886. The work at Northeastern University was supported by the US Department of Energy, Office of Science, Basic Energy Sciences contract number DE-FG02-07ER46352, and benefited from Northeastern University's Advanced Scientific Computation Center (ASCC), theory support at the Advanced Light Source, Berkeley and the allocation of time at the NERSC supercomputing center through DOE grant number DE-AC02-05CH11231.
\end{acknowledgement}

\begin{suppinfo}
Details of the computational model, dopant position determination, dopant distribution algorithm, and spectral gap correlations are given in this section.
\end{suppinfo}

\bibliography{BSCCO-dopants}

\end{document}